\documentclass[prl,twocolumn,draft,amsmath,showpacs]{revtex4}
\usepackage{graphics}

\begin{document}

\bibliographystyle{prsty}   
\input epsf

\title {Thermal and electrical transport in single crystalline MgB$_2$}

\author {A.V. Sologubenko, J. Jun, S. M. Kazakov, J. Karpinski, H.R. Ott  }
\affiliation{Laboratorium f\"ur Festk\"orperphysik, ETH H\"onggerberg,
CH-8093 Z\"urich, Switzerland}

\date{\today}

\begin{abstract}
We present the results of measurements of the thermal and electrical conductivity 
in the basal plane of single-crystalline  MgB$_{2}$ between 2 and 
300~K. The analysis of the 
temperature dependence of the thermal conductivity gives supporting 
evidence for two different gaps on different sheets of the Fermi 
surface in the superconducting state.  
The zero-temperature values of the two gaps are 
$\Delta_{1}(0)=5.3 \pm 0.5$~meV  and  $\Delta_{2}(0)=1.65 \pm 0.2$~meV.
\end{abstract}
\maketitle

The recent discovery of superconductivity in MgB$_{2}$ below an 
unexpectedly high
critical temperature $T_c$ of the order of 40~K initiated intensive 
studies of its physical properties \cite{Nagamatsu01}. 
Numerous results indicate that the 
superconducting state of  MgB$_{2}$ is well described by the original BCS theory. 
Most experiments  are compatible with a nodeless 
superconducting order parameter and imply that the electron-phonon interaction 
is responsible for
the pairing of the electrons \cite{Buzea01cm}. 
However, various types of experiments (see 
Ref.~\cite{Bouquet01cm} and references therein), mainly using 
powder or polycrystalline samples, have given 
evidence for two gaps in the quasiparticle excitation spectrum of this 
superconductor. Qualitative and especially quantitative experimental 
checks of this intriguing situation on single crystalline material seem in order.

Below we present results of measurements of the thermal conductivity 
$\kappa$ and the electrical resistivity $\rho$ parallel to the 
basal $ab$-plane of 
the hexagonal crystal lattice of  MgB$_2$ as a function of temperature 
$T$ between 2 and 300~K. 
We demonstrate that our low-temperature results cannot 
be explained in terms of a single-gap function below the critical 
temperature, but  are well approximated by assuming  two different energy 
gaps in the superconducting state. 

The investigated single crystal with dimensions of 
$0.5\times 0.17\times 0.035$ mm$^{3}$  was grown   with a high-pressure 
cubic anvil technique as described elsewhere \cite{Karpinski01}.  
The resistivity was measured by employing a 4-contact
configuration in the $ab$-plane. For the thermal conductivity measurements, 
the
standard method with a constant uniaxial heat flow was used. The same  
contacts were used for measuring the voltage and the temperature 
difference, respectively.  Additional  measurements of $\rho(T)$ 
and $\kappa(T)$ in constant magnetic 
fields $H$, oriented along the hexagonal $c$-axis, were made as 
well. 

\begin{figure}[t]
 \begin{center}
  \leavevmode
  \epsfxsize=0.9\columnwidth \epsfbox {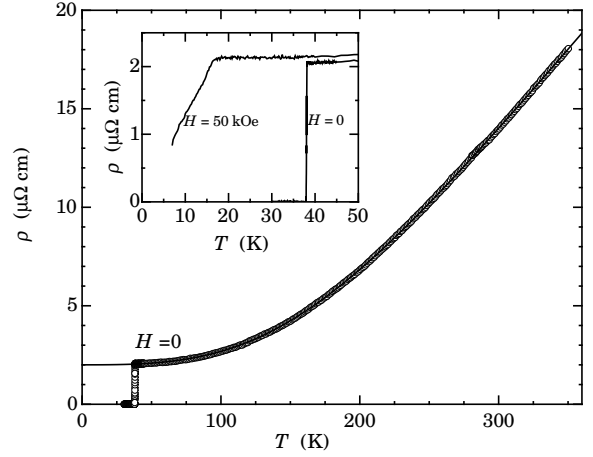}
   \caption{
  In-plane electrical resistivity $\rho(T)$ of hexagonal MgB$_{2}$. 
  The inset emphasizes the 
  low temperature part for $H=0$ and 50~kOe along the $c$-axis. 
  }
\label{R} 
\end{center}
\end{figure}

The electrical resistivity $\rho(T)$ in zero magnetic 
field is presented in Fig.~\ref{R}. 
The narrow ($\Delta T_{c}=0.1$~K) superconducting transition occurs at 
$T_c=38.1$~K. In the temperature region between $T_{c}$ and 130~K, $\rho(T)$ 
may be well approximated by $\rho = \rho_{0} + A' T^{3}$, 
where $\rho_{0}$ and $A'$ are constants. At temperatures below 50~K, 
$\rho \approx \rho_{0} = 2.0$~$\mu\Omega$~cm. 
A cubic $\rho(T)$ dependence has often been  observed in transition 
metals and is associated with interband electron-phonon scattering\cite{Mott35,Wilson38}. 
The corresponding equation 
\begin{equation}\label{eRho}
\rho (T)= \rho_0 + AT^3J_3\left( {\Theta_{R} / T} \right),
\end{equation}
where $J_n(z)=\int_0^z {x^ne^x\left( {1-e^x} \right)^{-2}dx}$,
fits our data  perfectly. 
This is demonstrated by the solid line in Fig.~\ref{R}. 
A similar 
fit to data from polycrystalline material is presented in 
Ref.~\cite{Putti01cm}.
The fit parameters $A$ and $\Theta_{R}$ characterize the strength of 
electron-phonon interaction and the cut-off frequency $\omega_{\rm max}$ 
of the phonon spectrum, respectively. 
The fit value of $\Theta_{R}=1226$~K is 
in excellent agreement with the value of $\hbar \omega_{\rm max} = 105$~meV 
measured in inelastic neutron scattering 
experiments \cite{Clementyev01,Muranaka01_INS}.
The magnitude of $\Theta_{R}$ is
considerably higher than the Debye 
temperature $\Theta_{D} \sim 700-900$~K, extracted from low-temperature specific heat 
measurements \cite{Budko01,Waelti01,Junod01cm}. 
This discrepancy is 
not unexpected  because electrical resistivity and specific heat involve 
different averages of the lattice frequencies \cite{Blackman51}. 

The inset of Fig.~\ref{R} emphasizes $\rho(T)$ close to 
the superconducting transitions in fields of $H=0$ and $H=50$~kOe. 
For $H=50$~kOe, the critical 
temperature,  marked by the onset of the deviation from the 
constant resistivity in the normal state, is shifted to $T_{c} = 17$~K. 

\begin{figure}[t]
 \begin{center}
  \leavevmode
  \epsfxsize=0.9\columnwidth \epsfbox {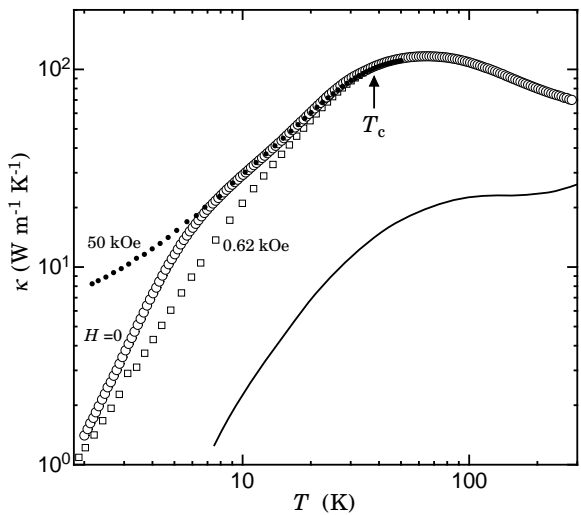}
   \caption{
  Thermal conductivity vs temperature in the $ab$-plane of MgB$_2$ 
  in zero magnetic field and for 
  $H=$ 0.62~kOe and 50~kOe parallel to the $c$-axis. 
  The solid line represents the typical 
  zero-field   behavior of $\kappa(T)$ measured on 
  polycrystalline samples (data from Ref.~\cite{Bauer01}). 
   }
\label{K} 
\end{center}
\end{figure}

The thermal conductivity data $\kappa(T)$ in zero magnetic field and 
for $H=$ 0.62 and 
50 kOe are presented in Fig.~\ref{K}. 
The zero-field $\kappa(T)$ values  are about an order of magnitude 
higher than previously reported for polycrystalline samples \cite{Muranaka01, 
Bauer01,Putti01cm,Schneider01}. Also the temperature dependence of 
$\kappa$ is quite different from those earlier data. 
Instead of a monotonous increase with 
temperature we note  a distinct maximum of 
$\kappa(T)$ at $T \sim 65$~K. 
No anomaly in $\kappa(T)$ provides evidence for the transition at 
$T_c$.
A pronounced slope change in $\kappa(T)$ 
is observed around $T \sim 6$~K, however. 

The thermal conduction of a superconductor is usually provided by 
electrons ($\kappa_e$) and phonons 
($\kappa_{\rm ph}$), such that 
\begin{equation}\label{eKappa}
\kappa = \kappa_e + \kappa_{\rm ph}.
\end{equation}
A separate identification of these two terms at arbitrary 
temperature is not straightforward. 
In the normal state, a convenient and often used way to estimate 
$\kappa_e$ is to employ the Wiedemann-Franz law, 
which relates the electrical resistivity and the electronic contribution to 
the thermal conductivity via 
\begin{equation}\label{eWFL}
\kappa_{e}(T) = L_{0} T / \rho(T),
\end{equation}
where $L_{0}=2.45\times 10^{-8}$ W~$\Omega$ K$^{-2}$ is the Lorenz number. 
The validity of this law 
requires an elastic scattering of electrons and 
it is well established that Eq.~(\ref{eWFL}) is applicable if the 
scattering of electrons by defects dominates. 
This is usually true at  low temperatures, 
where $\rho(T) \approx \rho_{0}$.  At higher temperatures but below 
$\Theta_{D}$, 
Eq.~(\ref{eWFL})
is usually invalid.
In our case, the  regime of 
applicability of Eq.~(\ref{eWFL}) is limited to temperatures below about 
50~K (see Fig.~\ref{R}). 
At temperatures $T_{c} \geq T \geq 50$~K,  $\kappa_{e}(T)$  
provides  about 
half of the total thermal conductivity. 

Upon decreasing the temperature below $T_{c}$, 
the reduction of the number of unpaired electrons 
leads to a decrease of  $\kappa_{e}$ and an increasing  $\kappa_{\rm ph}$. 
The 
overall behavior of $\kappa(T)$ in the superconducting state depends 
on the relative magnitudes of $\kappa_e$ and $\kappa_{\rm ph}$ and 
also on the strength of electron-phonon interaction.   
Using existing theoretical  models,
$\kappa(T)$ data may provide valuable 
information about the superconducting gap function. 
In the following, we restrict our analysis of $\kappa(T)$ to 
temperatures below 50~K, where the separation of $\kappa_{e}$ 
and $\kappa_{\rm ph}$ can reliably be accomplished. 

The electronic contribution to $\kappa(T)$ in the superconducting state 
has been calculated by 
Bardeen, Rickayzen, and Tewordt \cite{Bardeen59}. In their model
\begin{equation}\label{eKe}
\kappa_{e} = \kappa_{e,n} f(y),
\end{equation}
where
\begin{eqnarray}\label{eKes}
f(y) &=& {\frac{  2F_1(-y) + 2y\ln(1+e^{-y}) + \frac {y^2} {1+e^{y}} }   { 
2F_1(0)  } }, \quad(T<T_{c})\nonumber\\
{\rm and}&&\\
f(y) &=& 1,\quad(T \geq T_{c}),\nonumber
\end{eqnarray}
as well as $F_{n}(-y) = \int_0^\infty z^{n} (1+e^{z+y})^{-1} dz$, $y=\Delta(T)/k_{B} T$, 
and $\Delta(T)$ representing the energy gap. 
Here $\kappa_{e,n}$  is the normal-state electronic 
thermal conductivity and in our analysis $\kappa_{e,n}=L_{0} T / \rho_{0}$. 

The phonon thermal conductivity was analyzed in 
terms of the Debye-type relaxation rate approximation. The 
corresponding equation is \cite{BermanBook}
\begin{equation}\label{eKph}
    \kappa_{\rm ph} =  \frac{ k_{\rm B}}{ 2 {\pi}^{2} v } \left( 
    \frac{k_{\rm     B}}{\hbar} \right) ^{3} T^{3}  
    \int\limits_{0}^{\Theta_{D}/T}
    \frac{x^{4}e^{x}}{(e^{x}-1)^{2}}\tau (\omega,T) dx,
\end{equation}
where $\omega$ is the frequency of a phonon, $\tau(\omega,T)$ is the 
corresponding relaxation 
time, 
$ v=\Theta _D\left( {{{k_B} \mathord{\left/ {\vphantom {{k_B} \hbar }} \right. 
\kern-\nulldelimiterspace} \hbar }} \right)(6\pi ^2n)^{-1/3}$ 
is the average sound velocity,  $n$ is the number density of atoms, and $x=\hbar\omega/k_{\rm B} T$.  
For our analysis, we use $\Theta_{D}=750$~K, as deduced from 
specific heat measurements \cite{Waelti01,Budko01}.
The total phonon relaxation rate  
\begin{eqnarray}\label{eTau}
\tau^{-1} = L/v +   B \omega^{4} + C T \omega^{2} 
\exp(-\Theta_{D}/bT) + D \omega\nonumber\\
+ E \omega g(x,y)
\end{eqnarray}
can be represented as a 
sum of  terms  corresponding to independent scattering mechanisms.
The individual terms, dominating in different temperature intervals, 
introduce phonon scattering by sample boundaries, 
point defects, phonons, dislocations, and electrons,   respectively. 
The constants $L$, $B$, $C$,  $b$, $D$, and $E$ are a measure for the 
intensity 
of corresponding phonon relaxation processes. 
The function $g(x,y)$ is given in Ref.~\cite{Bardeen59}; we do 
not reproduce its rather complex form here. 

For $H=50$~kOe the phonon contribution $\kappa_{\rm ph}(T)$ in the normal state  (above 17~K), 
extracted from $\kappa(T)$ and  $\rho(T)$
using Eqs.~(\ref{eKappa}) and (\ref{eWFL}), is identical to
$\kappa_{\rm ph}(T)$ calculated for $H=0$ above $T_c$. 
We thus conclude that magnetic field effects on 
$\kappa_{\rm ph}$ are negligible. 
Based on the assumption that  phonon scattering by both defects 
and phonons is the same in the normal and in the superconducting state, 
we estimate the temperature region where the number of unpaired
electrons is too small to produce significant contributions to both phonon 
scattering and heat transport. 
Taking into account the lowest previously claimed value of 
$\Delta(0)=1.7$~meV \cite{Tsuda01},  
the electronic quasiparticles are effectively excluded from heat 
transport below 4~K.  

\begin{figure}[t]
 \begin{center}
  \leavevmode
  \epsfxsize=0.9\columnwidth \epsfbox {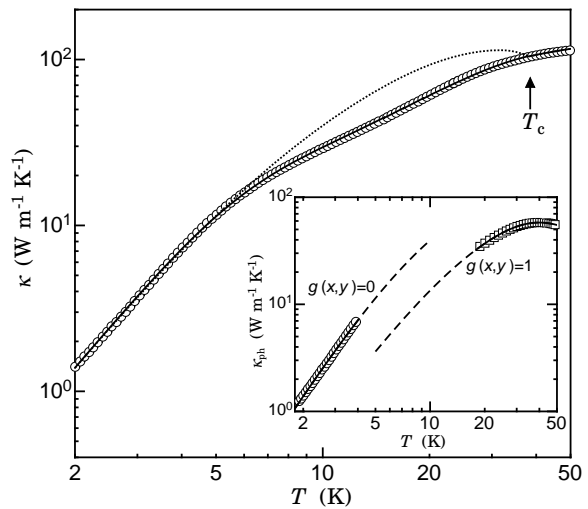}
   \caption{
  In-plane thermal conductivity vs temperature in zero magnetic field. 
  The solid line represents the best fit to the data assuming two gap 
  functions (see text). The dotted curve results from a standard single-gap 
  calculation 
  with $\Delta(0) = 1.76 k_{B}T_{c}$. 
  The inset represents the phonon thermal conductivity $\kappa_{\rm ph}$ in the regime 
  of negligible phonon-electron   scattering ($H=0$, open circles), 
  and the open squares represent $\kappa_{\rm ph}(T)$ at 
  temperatures where the electron-phonon scattering is that of the 
  normal-state ($H=50$~kOe, $T \geq T_{c}$).  
  The solid lines represent the fits described in the text.
  }
\label{KKfit} 
\end{center}
\end{figure}

The following  analysis procedure is based on 
the assumptions discussed above. 
In a first step,
the values of the parameters $L$, $B$, 
$C$,  $b$, $D$, and $E$, related to the  processes of 
phonon relaxation in the normal state, were obtained by fitting the 
$\kappa_{\rm ph}(T)$ data in $H = 50$~kOe
above 17~K (for this region $g(x,y)=1$) 
to Eqs.~(\ref{eKph}) and (\ref{eTau}) as well as
$\kappa(T)$ below 4~K where 
$g(x,y)=0$ and $\kappa_{e}=0$. 
These fits are shown in the inset of Fig.~\ref{KKfit}. 
The values of the parameters, common for both fits, are $L=2.6 \times 10^{-4}$~m, $B=1.07 
\times 10^{-3}$~s$^{3}$, $C=3.0\times 10^{-18}$~s~K$^{-1}$,  $b=6.2$, 
$D=2.1\times 10^{-5}$, and $E=7.5\times 10^{-5}$. 
We refrain from a detailed discussion of the parameters related to 
defect, boundary, and phonon-phonon scattering, and in the subsequent analysis
we consider them as serving as a background. 
Our primary interest is the   
electron-phonon interaction represented by the 
parameter $E$. 
The value of $E$ obviously fixes the shift between the two broken lines in the inset of 
Fig.~\ref{KKfit}. 
We note that the value of $E$ is quite robust with respect to the 
choice of different  phonon scattering terms in our fitting procedure.

In the essential step of the analysis, we aimed at establishing 
the temperature dependence of the gap function $\Delta(T)$, using Eq.~(\ref{eTau}).
The immediate result of this 
analysis is that a single gap function with the  temperature dependence  given 
by the BCS theory \cite{Muehlschlegel59} is not adequate for MgB$_{2}$. 
The significant reduction of the low-temperature thermal 
conductivity by a rather weak magnetic field of 0.62~kOe (see 
Fig.~\ref{K}) implies a strong interaction between 
phonons and the quasiparticles in the cores of the vortices induced by the 
magnetic field.
In case of such a  strong electron-phonon 
interaction and the formation of a single BSC-type gap,  
a significant change of the slope of $\kappa(T)$  
at $T_{c}$ is expected, as illustrated in 
Fig.~\ref{KKfit}, where the dotted line represents the 
calculation of $\kappa(T)$ for $\Delta(0)=1.76 k_{B}T_{c}$  and  $\Delta(T)$  
as tabulated in Ref.~\cite{Muehlschlegel59}. 
Hence,  for the majority of quasiparticles 
involved in phonon scattering  below $T_{c}$, the gap opens much more slowly 
with decreasing temperature
than expected 
from the standard BCS prediction. 
Our fitting procedure is much more successful if we assume the 
formation of two 
energy gaps, as has previously been claimed in reports of other experimental 
studies of MgB$_{2}$. Thus, we consider two subsystems of quasiparticles with 
gaps $\Delta_1$ and $\Delta_2$, different parameters $E_1$ and $E_2$ of 
phonon-electron scattering, and separate contributions 
$\kappa_{e1}$ and $\kappa_{e2}$ to the heat transport.
The normal-state parameter $E$ is then given by the sum $E = E_1 + E_2$. 
In the superconducting state, the last term in Eq.~(\ref{eTau})
adopts the form 
\begin{equation}\label{eEP2Gaps}
 E\omega \left[  (1-\alpha)  g(x,y_1)  +  \alpha g(x,y_2)  \right],
\end{equation}
where $y_1=\Delta_1(T)/k_B T$ and $y_2=\Delta_2(T)/k_B T$. The parameter 
$\alpha= E_2 / (E_1 + E_2)$ characterizes the relative 
weight of phonon scattering by quasiparticles in the two subsystems. 
Correspondingly,  
\begin{equation}\label{eKe2Gaps}
 \kappa_{e} = { \frac {L_0 T} {\rho_0}} \left[  (1-\delta) f(y_1) +  \delta f(y_2) \right],
\end{equation}
where the parameter $\delta= \kappa_{e2} / \kappa_{e}$ 
and $\kappa_{e} = \kappa_{e1} + \kappa_{e2}$. 
The fit is shown in Fig.~\ref{KKfit} as the solid line.

The resulting values of the fit parameters are 
$\Delta_{1}(0)=5.3 \pm 0.5$~meV,  $\Delta_{2}(0)=1.65 \pm 0.2$~meV,  
$\alpha = 1 \pm 0.02$ and 
$\delta=0.33 \pm 0.02$.  The ratio $2\Delta_{1}(0)/(k_{B}T_c)=3.22$ is close to 
the weak-coupling BCS value of 3.52. 
The second gap, however, is considerably smaller.
Our identification of two gaps is 
in qualitative and quantitative agreement with conclusions from results of specific heat, 
point-contact spectroscopy, and photoemission experiments  
(see, e.g., Table~I in Ref.~\cite{Bouquet01cm}).  
The two gaps in MgB$_{2}$ are,  by most authors, associated with 
different 
sheets of the Fermi-surface.  
According to density-functional calculations of Liu {\it et 
al.} \cite{Liu01}, the smaller and the larger gap are associated with 3-dimensional 
sheets and 2-dimensional tubes, respectively.  However, the 
first-principle FLAPW band calculations of Hase and Yamaji \cite{Hase01} 
claim the opposite.
Our result of $\alpha=1$ indicates that the dominant 
part of the interaction between quasiparticles and low-frequency phonons 
is provided by that part of the electronic excitation spectrum 
experiencing the smaller gap. 
The parameter $E$ in Eq.~(\ref{eTau}) 
can be expressed as
\begin{equation}\label{eE}
    E = \frac{m^{*2} E_{\rm def}^{2}} {2 \pi v \hbar^{3} D_0},
\end{equation}
where $m^{*}$ is an effective electron mass, $ E_{\rm def}$ is the 
deformation potential and $D_0$ is the mass density. 
The effective masses of quasiparticles on different parts of the Fermi surface of MgB$_{2}$ 
do not differ much \cite{Kong01cm}.
It has been shown 
\cite{An01} that the interaction between the holes of the cylindrical 
$\sigma$-bands and the in-plane  bond-stretching optical mode $E_{2g}$ with 
a frequency of 470~cm$^{-1}$ is provided by an 
ultrastrong deformation potential. 
If $E_{\rm def}$ is also strong for low-frequency acoustic phonons of the 
same polarisation, then
our result implies that it is the holes of the 2D sheets of the Fermi surface that 
experience the smaller energy gap, 
a counterintuitive result but in agreement with the conclusions of 
Ref.~\cite{Hase01}. 
Considering the contributions of the two subgroups of 
quasiparticles to the heat transport, our result  $\delta=0.33$ suggests
that the electrons experiencing the larger gap
provide approximately 2/3 of total electronic 
heat conduction. 

In conclusion, we present a study of the thermal 
conductivity $\kappa$ in the $ab$-plane of a single crystal of MgB$_{2}$. 
The analysis of the temperature dependence of $\kappa$
provides new convincing evidence for the formation of two superconducting gaps 
which are  associated with 
different sheets of the Fermi surface.  

\acknowledgments
Useful discussions with I.~L. Landau, R. Monnier and F. Bouquet are acknowledged.
This work was financially supported in part by
the Schweizerische Nationalfonds zur F\"{o}rderung der Wissenschaftlichen
Forschung.

\end{document}